# Enhancement of Noisy Speech exploiting a Gaussian Modeling based Threshold and a PDF Dependent Thresholding Function

Md Tauhidul Islam[a], Celia Shahnaz[b,*]

[a]*Department of Electrical and Computer Engineering, Texas A&M University, College Station, Texas, USA-77840*
[b]*Department of Electrical and Electronic Engineering, Bangladesh University of Engineering and Technology, Dhaka-1000, Bangladesh*

**Abstract**

This paper presents a speech enhancement method, where an adaptive threshold is statistically determined based on Gaussian modeling of Teager energy (TE) operated perceptual wavelet packet (PWP) coefficients of noisy speech. In order to obtain an enhanced speech, the threshold thus derived is applied upon the PWP coefficients by employing a Gaussian pdf dependent custom thresholding function, which is designed based on a combination of modified hard and semisoft thresholding functions. The effectiveness of the proposed method is evaluated for car and multi-talker babble noise corrupted speech signals through performing extensive simulations using the NOIZEUS database. The proposed method is found to outperform some of the state-of-the-art speech enhancement methods not only at at high but also at low levels of SNRs in the sense of standard objective measures and subjective evaluations including formal listening tests.

*Keywords:* Speech enhancement, perceptual wavelet packet transform, Teager energy, Gaussian PDF, Kullback-Liebler divergence

## 1. Introduction

Determination of a signal that is corrupted by additive or multiplicative noise has been of interest because of its importance in both theoretical and practical field. The main interest is to recover the real signal from the noise-mixed data received from microphone, ecg machine, radar, mobile phone or any other sound devices. Our aim is to make the recreated signal close to the original one. The use of such operation has application in broad area of speech communication applications, such as mobile telephony, speech coding and recognition, and hearing aid devices [8, 18, 21].

Over the decades, several methods have been developed to solve the noise reduction and speech enhancement problem. We can divide these methods in mainly three categories based on their domains of operation, namely time domain, frequency domain and time-frequency domain. Time domain methods include the subspace approach [11, 27], frequency domain methods include methods based on discrete cosine transform [5], spectral subtraction

[*]Corresponding author
*Email address:* celia.shahnaz@gmail.com (Celia Shahnaz)

[4, 26], minimum mean square error (MMSE) estimator [10, 17], wiener filtering [1, 3] and time frequency-domain methods involve the employment of the family of wavelets [2, 9, 12, 15, 25]. All these methods have their advantages and disadvantages. Time domain methods like subspace method provide a tradeoff between speech distortion and residual noise. But they burden a heavy computational load and as a result real-time processing becomes very difficult with these methods. On the other hand, frequency domain methods provide the advantage of real-time processing with less computation load. Among frequency domain methods, the most prominent one is spectral subtraction method. This method provides the facility of deducting noise from the noisy signal based on stationary nature of noise in speech signals. But this method has a major drawback of producing an artifact named musical noise which is perceptually disturbing, made of different tones of random frequencies and has an increasing variance. In the MMSE estimator, the spectral amplitude of noisy signal is modified based on the minimum square error. A large variance as well as worst performance in highly noisy situation are the main problems with this method. The main problem of wiener filter based methods is the necessity of clean speech statistics for their implementation. Like MMSE estimators, wiener filters also try to reach at an optimum solution depending on the error value between the computed signal with the real one.

The methods which use thresholding as process of removing noise, Universal threshold [9], SURE[20], WPF[2] and BayesShrink[6] are the prominent ones. In Universal thresholding method, a common threshold derived from noise power is used to threshold the wavelet coefficients. SURE applies Steins Uncertainty and BayesShrink applies Bayes principle to determine the threshold. WPF is modified version of Universal threshold method with speech and silent frame detection ability.

In this paper, we develop a speech enhancement method in the PWP domain, where the threshold is determined by performing the Gaussian statistical modeling of the TE operated PWP coefficients. Finally, a Gaussian pdf dependent custom thresholding function is employed on the PWP coefficients to obtain an enhanced speech. This thresholding function is designed based on the speech presence and absence probabilities so that it can perform thresholding operation in order to preserve the speech coefficients as well as to remove the noise coefficients.

The paper is organized as follows. Section II presents the Proposed Method. Section III describes results. Concluding remarks are presented in Section IV.

## 2. Proposed Method

The block diagram for the proposed method is shown in Fig. 1. It is seen from Fig. 1 that PWP transform is first applied to each input speech frame. Then, the PWP coefficients are subject to TE approximation with a view to determine a threshold value for performing thresholding operation in the WP domain. On using a custom thresholding function, an enhanced speech frame is obtained via inverse perceptual wavelet packet (IPWP) transform.

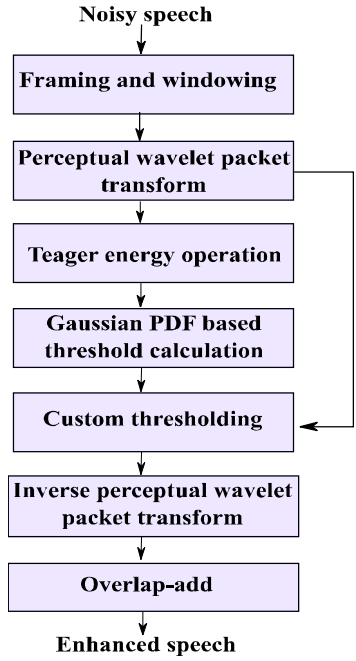

Figure 1: Block diagram for the proposed method

## 2.1. Perceptual Wavelet Packet Transform

The key element of the perceptual wavelet packet (PWP) transform is the use of the mel warping function to determine the WPT decomposition structure [24]. The main motivation behind using this transform is its ability to decompose the signal according to human auditory system. At low frequency, where human auditory system can differentiate the pitches precisely, PWP transform decomposes the signal in finer bands. On the other hand, at high frequency, PWP transform creates less number of bands as the human cochlea can not differentiate small differences in high frequency.

The perceptual mel scale is a scale of pitches judged by listeners to be equal in distance from one another. The conversion of frequency to mel is given in [24].

The clean and noise PWP coefficients in a subband of a noisy speech frame at an SNR of 5dB is plotted in Fig. 2(a). It is seen from this figure that for most of the coefficient indices, clean and noise PWP coefficients are not separable. Based on similar analysis performed on many speech signals corrupted by different noises, it is found that the time and frequency resolution provided by PWP transform is not sufficient to separate PWP coefficients of clean speech from that of noise even at a high SNR of 5dB. Since, TE operator has better time and frequency resolution [16], it can be very useful in handling noise. Therefore, we apply discrete time TE operator on the PWP coefficients.

*2.2. Teager Energy Operator*

Letting $W_{k,m}$ as the *m*-th PWP coefficient in the *k*-th subband, the *m*-th TE operated coefficient $t_{k,m}$ corresponding to the *k*-th subband of the PWP transform is given by

$$t_{k,m} = T[W_{k,m}]. \tag{1}$$

Fig. 2(b) presents the clean and noise TE operated PWP coefficients in a subband of a noisy speech frame at the same SNR as used in Fig. 2(a). It is seen from this figure that at the indices where TE operated PWP coefficients of clean speech have higher values, the TE operated PWP coefficients of noise show lower values. As a result, thresholding operation on the noisy PWP coefficients needs a low threshold value thus removing the noise leaving the speech undistorted. On the contrary, at the indices, where TE operated PWP coefficients of clean speech have lower values, the TE operated PWP coefficients of noise exhibit higher values as expected. Thus thresholding the noisy speech PWP coefficients needs a higher threshold value and removes the necessary noise without speech distortion at a significant level. Therefore, TE operation on PWP coefficients is found as more capable of serving the goal of thresholding operation by reducing the noise as well as preserving the speech.

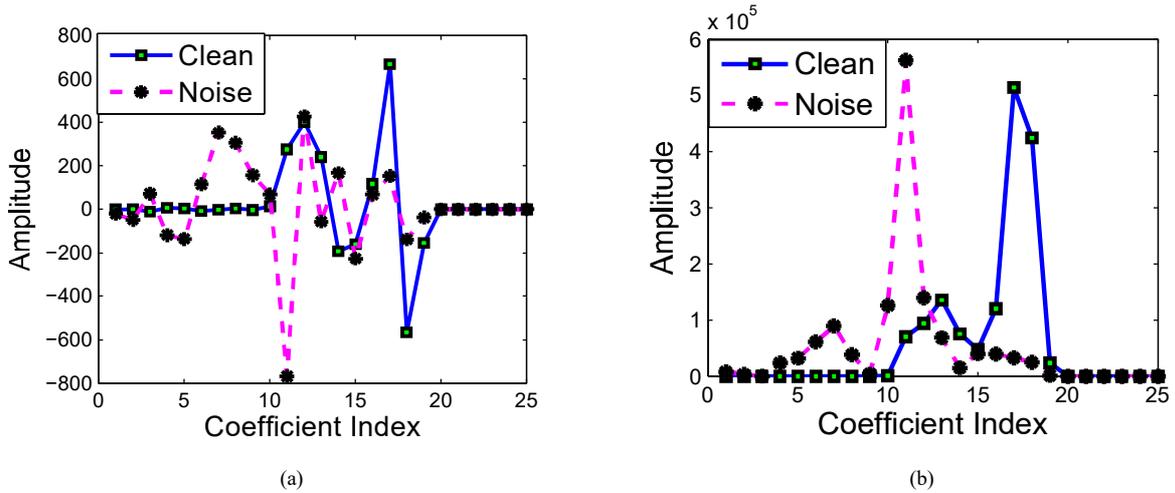

Figure 2: (a) PWP Coefficients (b) TE Operated PWP Coefficients of a noisy speech subband at an SNR of 5dB

*2.3. Proposed Model for TE Operated PWP Coefficients*

The outcome of a speech enhancement method based on the thresholding in a transform domain depends mainly on two factors, namely the threshold value and the thresholding functions. The use of a unique threshold for all the PWP subbands is not reasonable. As a crucial parameter, the threshold value in each subband is required to be adjusted very precisely so that it can prevent distortion in the enhanced speech as well as decrease annoying residual noise. By considering the probability distributions of the $t_{k,m}$ of the noisy speech, noise and clean speech, a more accurate threshold value can be obtained using a suitable pattern matching scheme or similarity measure. Since speech is a time-varying signal, it is difficult to realize the actual probability distribution function (pdf) of speech or its $t_{k,m}$. As an alternative to formulate a pdf of the of speech, we can easily formulate the histogram of its $t_{k,m}$ and can approximate the histogram by a reasonably close pdf namely Gaussian distribution. For the $t_{k,m}$s in a subband of a noisy speech frame, the empirical histogram along with the Gaussian distributions are superimposed in Fig. 3, 4 and 5 in presence of car noise at SNRs of −15, 0 and 15 dB. From this figure, it is obvious that Gaussian distribution fits the empirical histogram very finely. Similar analysis results are obtained for empirical histogram and Gaussian distribution of TE operated noise PWP coefficients at the same SNRs as used in Fig. 3, 4 and 5 and are shown in Fig. 6, 7 and 8.

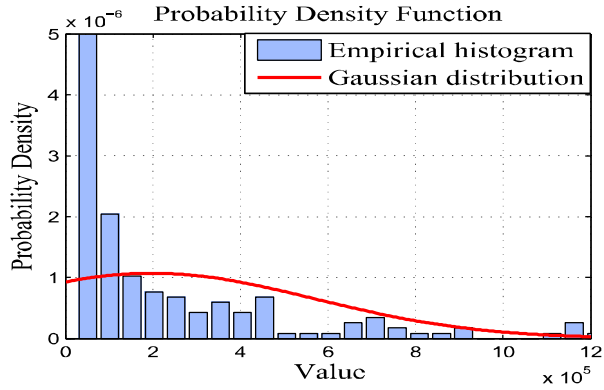

Figure 3: Empirical histogram and Gaussian distribution of TE operated PWP coefficients of noisy speech at SNR of −15 dB

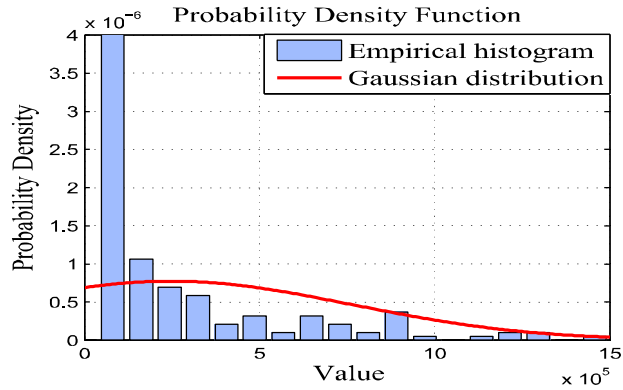

Figure 4: Empirical histogram and Gaussian distribution of TE operated PWP coefficients of noisy speech at SNR of 0 dB

The efficiency of the fitting of the data with the Gaussian distribution can also be proved through objective index Normalized Mean Square Error (NMSE), which is calculated by (2).

$$NMSE = \sqrt{\frac{1}{N}\sum_{i=1}^{N}(\frac{y_i - x_i}{y_i})^2}, \qquad (2)$$

where $y_i$ is the Gaussian fitted data, $x_i$ is the empirical data and $N$ is the total number of data points.

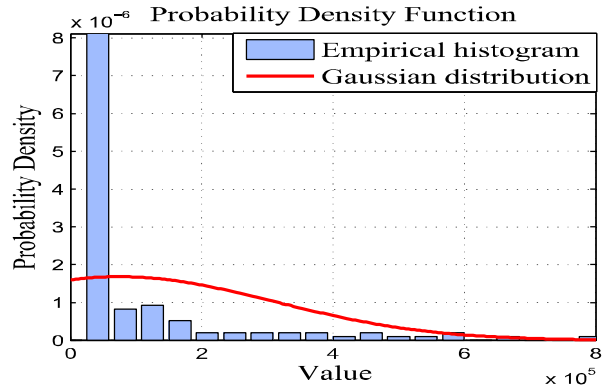

Figure 5: Empirical histogram and Gaussian distribution of TE operated PWP coefficients of noisy speech at SNR of 15 dB

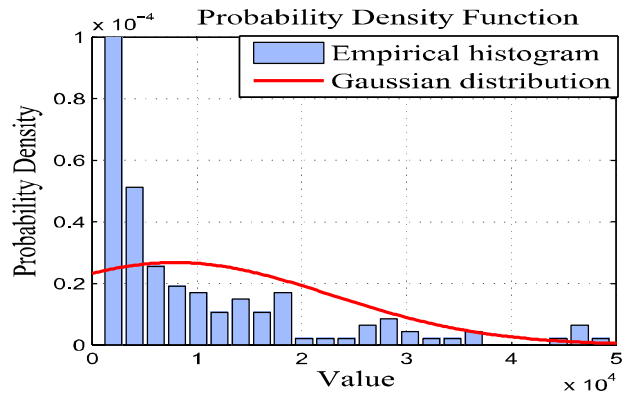

Figure 6: Empirical histogram and Gaussian distribution of TE operated noise PWP coefficients at SNR of −15 dB

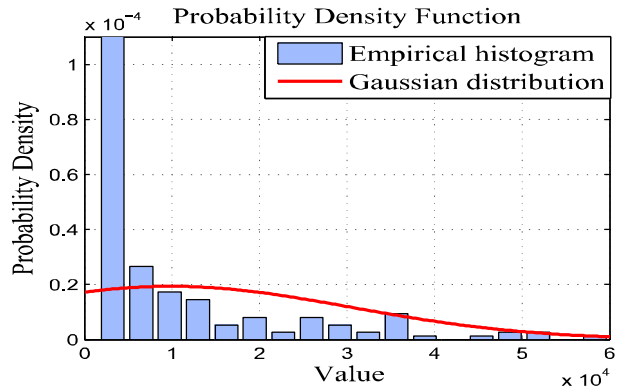

Figure 7: Empirical histogram and Gaussian distribution of TE operated noise PWP coefficients at SNR of 0 dB

The NMSE for data corrupted with car noise at different noise levels is plotted in Fig. 9.

It is seen from Fig. 9 that although for higher noise conditions like −15dB and −10dB, the NMSE is around 10%, for lower noise conditions, it is less than 8%, which attests that the Gaussian distribution finely fits the data.

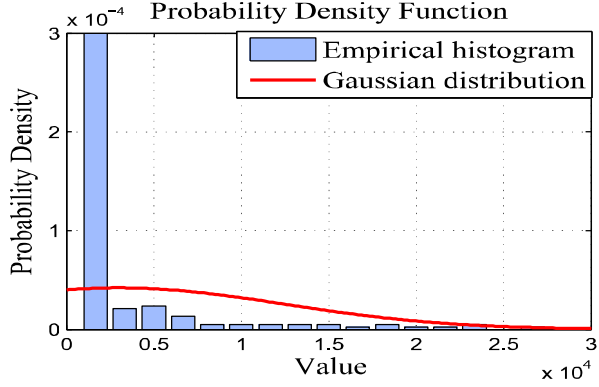

Figure 8: Empirical histogram and Gaussian distribution of TE operated noise PWP coefficients at SNR of 15 dB

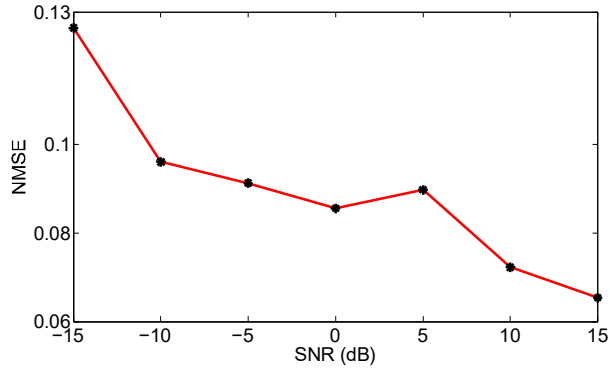

Figure 9: NMSE for data corrupted with car noise

*2.4. Determination of Proposed Adaptive Threshold*

The entropy of each subband of the PWP coefficients is found different from each other. So, an entropy measure may be chosen to select a suitable threshold value adaptive to each subband. Some popular similarity measures that are related to the entropy functions are the Variational distance, the Bhattacharyya distance, the Harmonic mean, the Kullback Leibler(K-L) divergence, and the Symmetric K-L divergence. The K-L divergence is always nonnegative and zero if and only if the approximate Gaussian distribution functions of the pdf of noisy speech and that of the noise or the approximate Gaussian distribution functions of the pdf of the noisy speech and that of the clean speech are exactly the same. In order to have a symmetric distance between the any two approximate Gaussian distribution functions as mentioned above, the Symmetric K-L divergence has been adopted in this paper. The Symmetric K-L divergence is defined as

$$SKL(p,q) = \frac{KL(p,q) + KL(q,p)}{2}, \quad (3)$$

where $p$ and $q$ are the two approximate Gaussian pdfs calculated from the corresponding histograms each having $N$ number of bins and $KL(p,q)$ is the K-L divergence given by

$$KL(p,q) = \sum_{i=1}^{N} p_i(t_{k,m}) ln \frac{p_i(t_{k,m})}{q_i(t_{k,m})}. \tag{4}$$

In (4), $p_i(t_{k,m})$ is the pdf of $t_{k,m}$ of noisy speech given by

$$p_i(t_{k,m}) = \frac{n_i}{N_c}, \tag{5}$$

where $n_i$ is number of coefficients in $i$-th bin and $N_c$ total number of coefficients in each subband. Similarly, the approximate Gaussian pdf of the $t_{k,m}$ of the noise and that of the $t_{k,m}$ of the clean speech can be estimated following (5) and denoted by $q_i(t_{k,m})$ and $r_i(t_{k,m})$, respectively. Below a certain value of threshold $\lambda$, the symmetric K-L divergence between $p_i(t_{k,m})$ and $q_i(t_{k,m})$ is approximately zero, i.e.,

$$SKL(p_i(t_{k,m}), q_i(t_{k,m})) \approx 0. \tag{6}$$

By solving the above equation, we get a value of $\lambda$ following [23],

$$\lambda(k) = \frac{\sigma_n(k)}{\sqrt{\gamma_k}} \sqrt{2(\gamma_k + \gamma_k^2)} \times ln(\sqrt{1 + \frac{1}{\gamma_k}}), \tag{7}$$

where $\gamma_k$ is segmental SNR of subband $k$ defined as

$$\gamma_k = \frac{\sigma_r^2(k)}{\sigma_n^2(k)}. \tag{8}$$

In this equation, $\sigma_r^2(k)$ is the signal power at $k$ subband and $\sigma_n^2(k)$ is the noise power at $k$ subband.

*2.5. Proposed Thresholding Function*

We propose a pdf dependent custom thresholding function derived from the modified hard and the semisoft thresholding functions [22]. Representing $\lambda(k)$ derived from (7) as $\lambda_1(k)$ and letting $\lambda_2(k) = 2\lambda_1(k)$, the proposed thresholding function is developed as

$$(Y_{k,m})_{PCT} = \begin{cases} \alpha(k,m) sgn(Y_{k,m}) \times G, & \text{if } |(Y_{k,m})| < \lambda_1(k) \\ Y_{k,m}, & \text{if } |(Y_{k,m})| > \lambda_2(k), \\ (1 - \alpha(k,m))\Pi_1 + \alpha(k,m)\Pi_2, & \text{otherwise,} \end{cases} \tag{9}$$

where

$$G = \frac{|(Y_{k,m})|^{\beta(k,m)}}{[\lambda_1(k)]^{(\beta(k,m)-1)}}, \tag{10}$$

$$\Pi_1 = sgn(Y_{k,m}) \times \lambda_2(k) \frac{|(Y_{k,m})| - \lambda_1(k)}{\lambda_2(k) - \lambda_1(k)}, \tag{11}$$

$$\Pi_2 = Y_{k,m}. \qquad (12)$$

In (9), $(Y_{k,m})_{PCT}$ stands for the PWP coefficients thresholded by the proposed custom thresholding function expressed from (9)-(12) and two shape parameters of the proposed thresholding function are represented by $\alpha(k,m)$ and $\beta(k,m)$.

The comparison of the proposed custom thresholding function with the conventional modified hard and semisoft thresholding functions is shown in Fig. 10. In the region between $\lambda_1$ and $\lambda_2$, this figure demonstrates the flexibility of the proposed thresholding operation in a sense that it can be viewed as $(1 - \alpha(k,m))(Y_{k,m})_{SS} + \alpha(k,m)(Y_{k,m})_{MH}$ which is a linear combination of the modified hard and the semisoft thresholding function. Here, $(Y_{k,m})_{MH}$ stands for the PWP coefficients thresholded by the modified hard thresholding function and $(Y_{k,m})_{SS}$ represents the PWP coefficients thresholded by the semisoft thresholding function. Unlike these functions, depending on the value of shape parameter $\alpha(k,m)$, it can be verified from (9) that the proposed thresholding function gets the following forms,

$$\lim_{\alpha(k,m) \to 0} (Y_{k,m})_{PCT} = (Y_{k,m})_{SS},$$

$$\lim_{\alpha(k,m) \to 1} (Y_{k,m})_{PCT} = (Y_{k,m})_{MH}.$$

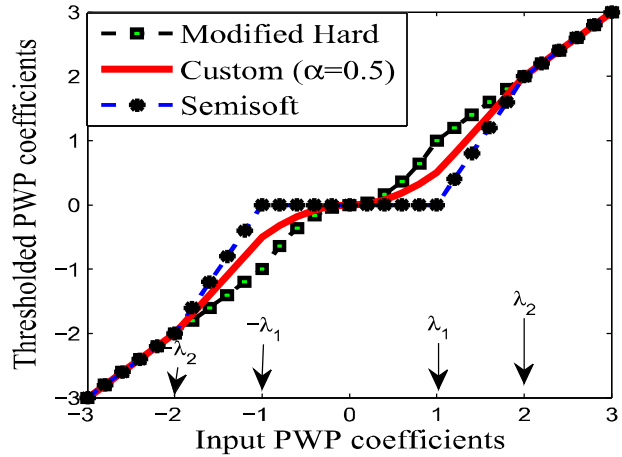

Figure 10: Input Output Relation for semisoft, modified hard and proposed custom thresholding function

*2.5.1. Effect of the Shape Parameters on the Proposed Thresholding Function*

In order to realize the effect of $\alpha(k,m)$ and $\beta(k,m)$ on the proposed thresholding function, the variation of $\alpha(k,m)$ and $\beta(k,m)$ for different values of $\frac{R(k,m)}{Q(k,m)}$ are obtained using (13) and (14) and plotted in Fig.11. From this figure, it is seen that for a large value of $\frac{R(k,m)}{Q(k,m)}$, $\alpha(k,m)$ becomes high, i.e., close to 1 that increases the probability of $Y_{k,m}$ to be a speech coefficient. In this case, $(Y_{k,m})_{PCT}$ acts like $(Y_{k,m})_{MH}$ as expected, since if a coefficient has a high probability to be speech should not be a thresholded to zero before $\lambda_1(k)$ and should be unchanged after $\lambda_1(k)$ as done in modified hard thresholding function. It is also found from Fig. 11 that for a small value of $\frac{R(k,m)}{Q(k,m)}$, $\alpha(k,m)$ becomes low, i.e., close to zero and $(Y_{k,m})_{PCT} \approx (Y_{k,m})_{SS}$. It is also expected since if the probability of a PWP coefficient to be speech becomes low, it should be thresholded to zero before $\lambda_1(k)$ and thresholded to a small value upto $\lambda_2(k)$ as done in semisoft thresholding function. From Fig.11, it is seen that for a small value of $\frac{R(k,m)}{Q(k,m)}$, $\beta(k,m)$ gets a high value that increases the probability of $Y_{k,m}$ to be a noise coefficient. In this case, it can be seen from (10) that $(Y_{k,m})_{PCT}$ in (9) tends to zero as expected since a noise PWP coefficient should be made zero to completely remove the noise. On the other hand, for a high value of $\frac{R(k,m)}{Q(k,m)}$, $\beta(k,m)$ becomes low that decreases the probability of $Y_{k,m}$ to be a noise coefficient. Therefore, from (10) and (9), it can be verified that $(Y_{k,m})_{PCT}$ gets a small value instead of being thresholded to zero. This is also expected since a PWP coefficient that has a less probability to be a noise coefficient should not be thresholded to zero.

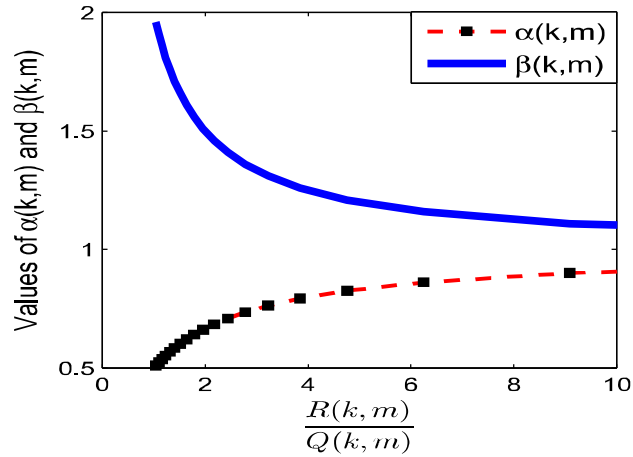

Figure 11: Plot of $\alpha(k,m)$ and $\beta(k,m)$ for different values of $\frac{R(k,m)}{q(k,m)}$

### 2.5.2. Determination of Shape Parameters

The proposed thresholding function can be adapted to noise characteristics of the input noisy speech based on the shape parameters $\alpha(k,m)$ and $\beta(k,m)$ which are defined as

$$\alpha(k,m) = \frac{1 + R(k,m)}{2(1 + Q(k,m))}, \quad (13)$$

$$\beta(k,m) = \frac{2(1 + Q(k,m))}{(1 + R(k,m))}, \quad (14)$$

where $R(k,m)$ and $Q(k,m)$ are the speech presence and absence probabilities, respectively, of the $m$-th coefficient in the $k$-th subband.

Given two hypotheses, $H_0$ and $H_1$, which indicate respectively speech absence and presence in the $m$-th coefficient of the $k$-th subband, and assuming a complex Gaussian distributions for both speech and noise PWP coefficients [10], the conditional pdfs of the speech and noise PWP coefficients are given by

$$f(Y(k,m)|H_0(k,m)) = \frac{1}{\pi\sigma_n^2} exp(-\frac{|Y(k,m)|^2}{\sigma_n^2}), \qquad (15)$$

$$f(Y(k,m)|H_1(k,m)) = \frac{1}{\pi(\sigma_n^2 + \sigma_r^2)} exp(-\frac{|Y(k,m)|^2}{\sigma_n^2 + \sigma_r^2}). \qquad (16)$$

Using *aposteriori* and *apriori* SNRs defined by [10]

$$\Upsilon(k,m) = \frac{|Y(k,m)|^2}{\sigma_n^2(k,m)}, \qquad (17)$$

$$\eta(k,m) = \frac{\sigma_r^2(k,m)}{\sigma_n^2(k,m)}, \qquad (18)$$

and following (15) and (16), the conditional pdfs of the *aposteriori* SNR can be written as [7]

$$f(\Upsilon(k,m)|H_0(k,m)) = e^{-\Upsilon(k,m)} I_2, \qquad (19)$$

$$f(\Upsilon(k,m)|H_1(k,m)) = \frac{1}{1+\eta(k,m)} \times exp(-\frac{\Upsilon(k,m)}{1+\eta(k,m)}) I_2. \qquad (20)$$

In (19) and (20), $I_2 = u(\Upsilon(k,m))$ is the unit step function. Noting that the conditional speech presence probability $R(k,m) = P(H_1(k,m)|\Upsilon(k,m))$, applying Bayes rule and using (20), an expression for $R(k,m)$ can be derived as

$$R(k,m) = [1 + \frac{Q(k,m)}{1-Q(k,m)}(1+\widehat{\eta}(k,m))exp(-v(k,m))]^{-1}, \qquad (21)$$

where $\widehat{\eta}(k,m)$ is the estimated *apriori* SNR obtained as in [7] and

$$v(k,m) = \frac{\widehat{\eta}(k,m)\Upsilon(k,m)}{(1+\widehat{\eta}(k,m))}. \qquad (22)$$

Speech absence probability $Q(k,m)$ in (21) can be determined as

$$Q(k,m) = 1 - R_{local}(k,m) R_{global}(k,m) R_{subband}(k,m). \qquad (23)$$

In (23), $R_{local}(k,m)$ and $R_{global}(k,m)$ are the speech presence probabilities in local and global windows in the PWP domain. Letting $\tau$ for representing either "local" or "global" window, $R_\tau(k,m)$ can be given by

$$R_\tau(k,m) = \begin{cases} 0, & \text{if } \xi_\tau(k,m) \leq \xi_{min} \\ 1, & \xi_\tau(k,m) \geq \xi_{max}, \\ \frac{log(\xi_\tau(k,m)/\xi_{min})}{log(\xi_{max}/\xi_{min})}, & \text{otherwise,} \end{cases} \qquad (24)$$

where $\xi_\tau(k,m)$ representing either "local" or "global" average of the *apriori* SNR given by

$$\xi_\tau(k,m) = \sum_{i=-W_\tau}^{i=W_\tau} h_\tau(i)\xi(k-i,m). \tag{25}$$

In (25), $h_\tau$ is a normalized window of size $2w_\tau + 1$ and $\xi(k,m)$ represents a recursive average of the *apriori* SNR given by

$$\xi(k,m) = \kappa\xi(k,m-1) + (1-\kappa)\widehat{\eta}(k,m-1), \tag{26}$$

where $\kappa$ denotes a smoothing constant. Note that in (24), $\xi_{min}$ and $\xi_{max}$ are the two empirical constants representing minimum and maximum values of $\xi(k,m)$ given in (26). $R_{subband}(k)$ in (23) can be computed as

$$R_{subband}(k) = \begin{cases} 0, & \text{if } \xi_{subband}(k) < \xi_{min} \\ 1, & \text{if } \xi_{subband}(k) > \xi_{subband}(k-1) \text{ and } \xi_{subband}(k) > \xi_{min}, \\ \mu(k), & \text{otherwise,} \end{cases} \tag{27}$$

where $\mu(k)$ is expressed as

$$\mu(k) = \begin{cases} 0, & \text{if } \xi_{subband}(k) \leq \xi_{peak}(k)\xi_{min} \\ 1, & \text{if } \xi_{subband}(k) \geq \xi_{peak}(k)\xi_{max}, \\ \frac{log(\xi_{subband}(k)/\xi_{peak}(k)/\xi_{min})}{log(\xi_{max}/\xi_{min})}, & \text{otherwise.} \end{cases} \tag{28}$$

In (28) and (27), $\xi_{subband}(k)$ is determined as

$$\xi_{subband}(k) = \frac{1}{N_c} \sum_{1 \ll m \ll N_c} \xi(k,m) \tag{29}$$

and $\xi_{peak}$ in (27) is a confined peak value of $\xi_{subband}(k)$. Thus computing $R(k,m)$ and $Q(k,m)$ following (21) and (23), the shape parameters $\alpha(k,m)$ and $\beta(k,m)$ can be determined using (13) and (14), respectively.

*2.6. Inverse Perceptual Wavelet Packet Transform*

For a noisy speech frame, we obtain thresholded PWP coefficients using the proposed threshold in (7) and the proposed thresholding function in (9). An enhanced speech frame $\widehat{r}[n]$ is synthesized by performing inverse PWP transform as

$\widehat{r}[n] = PWP^{-1}(Y_{k,m})_{PCT}.$

The enhanced speech signal is reconstructed by using the standard overlap-and-add method [21].

**3. Results**

In this Section, a number of simulations is carried out to evaluate the performance of the proposed method.

Table 1: Constants used to determine the shape parameters

| Constants | Value |
|---|---|
| $\beta$ | 0.7 |
| $\xi_{min}$ | -10 dB |
| $\xi_{max}$ | -5 dB |
| $\xi_{peak}$ | 10 dB |
| $w_{local}$ | 1 |
| $w_{global}$ | 15 |

*3.1. Simulation Conditions*

Real speech sentences from the NOIZEUS database are employed for the experiments, where the speech data is sampled at 8 KHz [13]. To imitate a noisy environment, noise sequence is added to the clean speech samples at different SNR levels ranging from 15 dB to -15 dB. As in [19], two different types of noises, such as car and babble are adopted from the NOIZEUS databases [13].

In order to obtain overlapping analysis frames, hamming windowing operation is performed, where the size of each of the frame is 512 samples with 50% overlap between successive frames. A 6-level PWP decomposition tree with 10 db bases function is applied on the noisy speech frames [24], [23] resulting in subbands $k = 1, 2, .....24$.

The values of used constants to determine the shape parameters in the proposed thresholding function are given in table 1.

*3.2. Comparison Metrics*

Standard Objective metrics namely, Segmental SNR (SNRSeg) improvement in dB, Perceptual Evaluation of Speech Quality (PESQ) and Weighted Spectral Slope (WSS) are used for the evaluation of the proposed method [18]. The proposed method is subjectively evaluated in terms of the spectrogram representations of the clean speech, noisy speech and enhanced speech. Formal listening tests are also carried out in order to find the analogy between the objective metrics and the subjective sound quality. The performance of our method is compared with some of the state-of-the-art speech enhancement methods, such as Universal [9], SMPO [19] and Wavelet Packet Transform based method with Modified Hard Thresholding Function (WPMH) [23] in both objective and subjective senses.

*3.3. Objective Evaluation*

*3.3.1. Results for Speech signals with Car Noise*

SNRSeg improvement, PESQ and WSS for speech signals corrupted with car noise for Universal, SMPO, WPCT and proposed methods are shown in Fig.12, Table 2 and Fig.13.

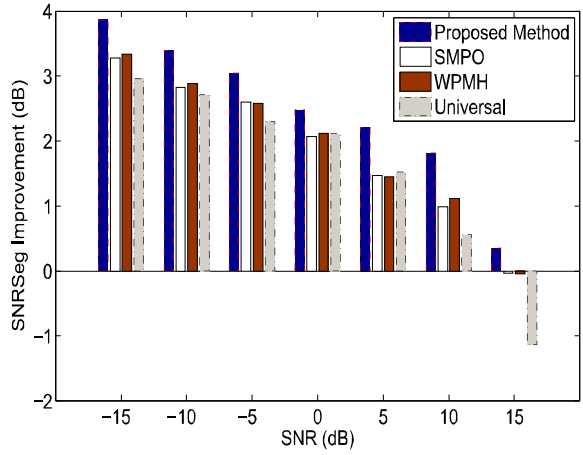

Figure 12: SNRSeg Improvement for different methods in car noise

Table 2: PESQ for different methods in car noise

| SNR(dB) | Universal | SMPO | WPCT | Proposed Method |
|---|---|---|---|---|
| -15 | 1.16 | 1.15 | 1.19 | 1.27 |
| -10 | 1.23 | 1.37 | 1.36 | 1.45 |
| -5 | 1.32 | 1.51 | 1.57 | 1.61 |
| 0 | 1.43 | 1.69 | 1.80 | 1.79 |
| 5 | 1.69 | 2.07 | 2.11 | 2.13 |
| 10 | 1.93 | 2.38 | 2.45 | 2.43 |
| 15 | 2.14 | 2.60 | 2.69 | 2.75 |

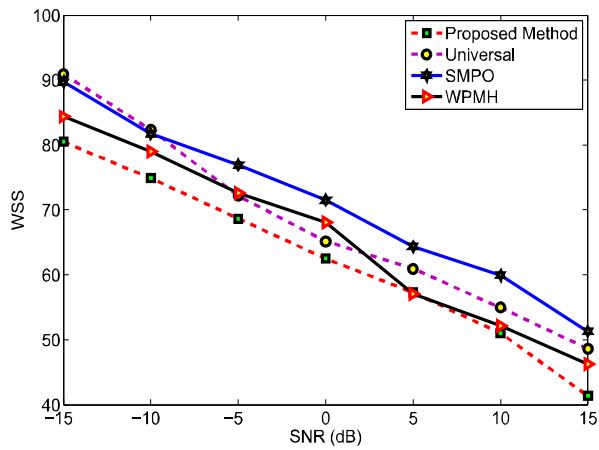

Figure 13: WSS for different methods in car Noise

In Fig.12, the performance of the proposed method is compared with that of the other methods at different levels of SNR for car noise in terms of Segmental SNR improvement. We see, the SNRSeg improvement increases as SNR decreases. At a low SNR of $-15dB$, the proposed method yields the highest SNRSeg improvement. Such larger values of SNRSeg improvement at a low level of SNR attest the capability of the proposed method in producing enhanced speech with better quality for speech severely corrupted by car noise.

In Table 2, it can be seen that at a low level of SNR, such as $-15dB$, all the methods show lower values of PESQ scores, whereas the PESQ score is much higher, as expected, for the proposed method. The proposed method also yields larger PESQ scores compared to that of the other methods at higher levels of SNR. Since, at a particular SNR, a higher PESQ score indicates a better speech quality, the proposed method is indeed better in performance in the presence of a car noise.

Fig.13 represents the WSS values as a function of SNR for the proposed method and that for the other methods. As shown in the figure, the WSS values resulting from all other methods are relatively larger for a wide range of SNR levels, whereas the proposed method is capable of producing enhanced speech with better quality as it gives lower values of WSS even at a low SNR of $-15dB$.

*3.3.2. Results for Speech signals with Multi-talker Babble Noise*

SNRSeg improvement, PESQ and WSS for speech signals corrupted with babble noise for Universal, SMPO and proposed methods are shown in Fig.14, 16 and 15, respectively.

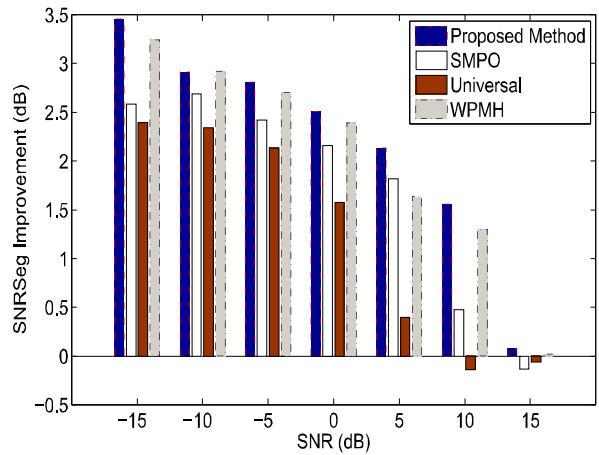

Figure 14: SNRSeg Improvement for different methods in babble noise

In Fig. 14, it can be seen that at a low level of SNR of $-15dB$, the proposed method provides a SNRSeg improvement that is significantly higher than that of the methods of comparison. The proposed method still shows better performance in terms of SNRSeg improvement for higher SNRs also.

For speech corrupted with babble noise, in Fig.15, the mean values of PESQ with standard deviation obtained using the proposed method is plotted and compared with that of the other methods. From this plot, it is seen that over the whole SNR range considered, the proposed method continue to provide higher PESQ with almost non-overlapping standard deviation in the presence of babble noise.

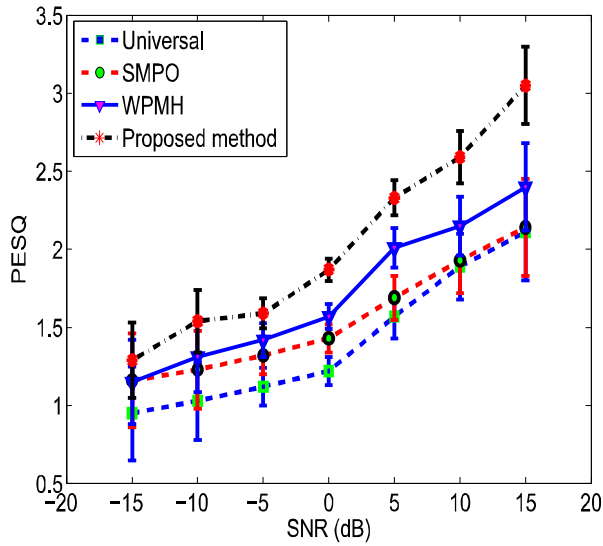

Figure 15: PESQ for different methods in babble noise

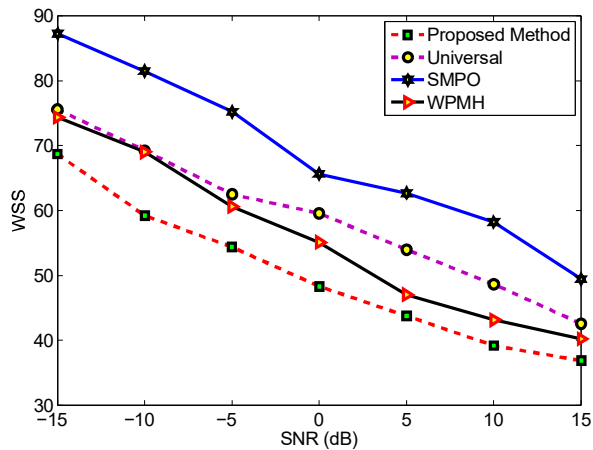

Figure 16: WSS for different methods in babble noise

The performance of the proposed method is compared with that of the other methods in terms of WSS in Fig.16 at different levels of SNRs in presence of babble noise. It is clearly seen from this figure that WSS increases as SNR decreases. At a low SNR of $-15dB$, the proposed method yields a WSS that is significantly lower than that of all other methods, which remains lower over the higher SNRs also.

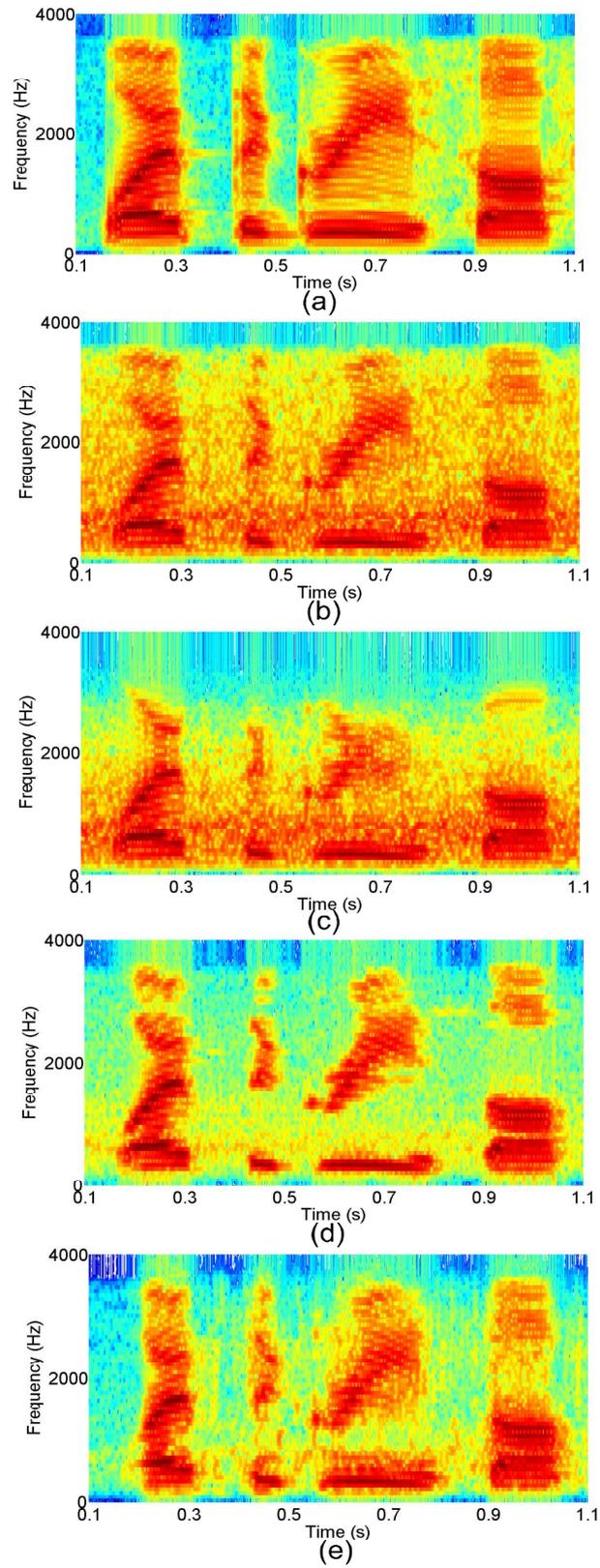

Figure 17: Spectrograms of (a) Clean Signal (b) Noisy Signal with 10dB car noise; spectrograms of enhanced speech from (c) Universal method (d) SMPO method (e) Proposed Method

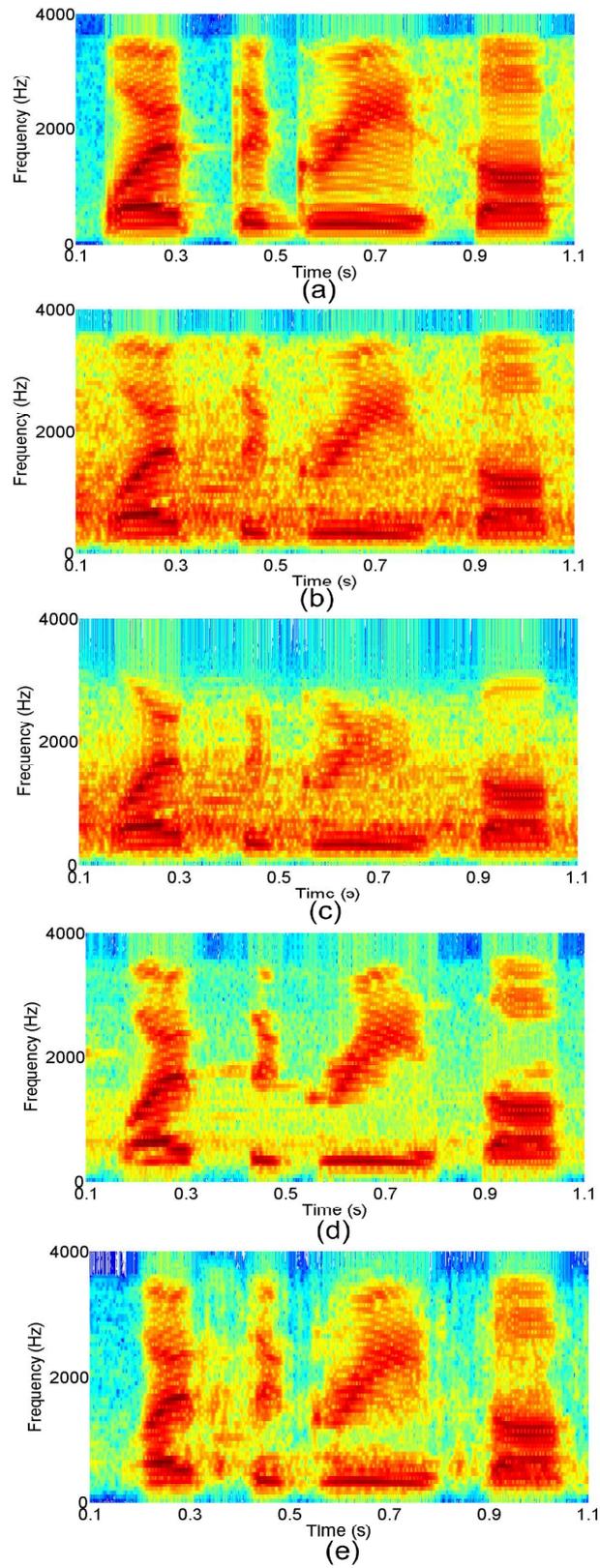

Figure 18: Spectrograms of (a) Clean Signal (b) Noisy Signal with 10dB babble noise; spectrograms of enhanced speech from (c) Universal method (d) SMPO method (e) Proposed Method

*3.4. Subjective Evaluation*

In order to evaluate the subjective observation of the enhanced speech, spectrograms of the clean speech, the noisy speech, and the enhanced speech signals obtained by using the proposed method and all other methods are presented in Fig. 17 for car noise corrupted speech at an SNR of 10 dB. It is evident from this figure that the harmonics are well preserved and the amount of distortion is greatly reduced in the proposed method. Thus, the spectrogram observations with lower distortion also validate our claim of better speech quality as obtained in our objective evaluations in terms of higher SNR improvement in dB, higher PESQ score and lower WSS in comparison to the other methods. Another set of spectrograms for babble noise corrupted speech at an SNR of 10 dB is also presented in Fig.18. This figure attests that the proposed method has a better efficacy in preserving speech harmonics even in case of babble noise.

Formal listening tests are also conducted, where ten listeners are allowed and arranged to perceptually evaluate the enhanced speech signals. A full set (thirty sentences) of the NOIZEUS corpus was processed by Universal, SMPO and proposed method for subjective evaluation at different SNRs. Subjective tests were performed according to ITU-T recommendation P.835 [13]. In this tests, a listener is instructed to successively attend and rate the enhanced speech signal based on (a) the speech signal alone using a scale of SIG (1 = very unnatural, 5 = very natural), (b) the background noise alone using a scale of background conspicuous/ intrusiveness (BAK) (1 = very conspicuous, very intrusive; 5 = not noticeable), and (c) the overall effect using the scale of the mean opinion score (OVRL) (1 = bad, 5 = excellent). More details about the testing methodology can be found in [14]. The mean scores of SIG, BAK, and OVRL scales for the three speech enhancement methods evaluated in the presence of car noise at an SNR of 5 dB are shown in Tables 3, 4, and 5. For the three methods evaluated using babble noise-corrupted speech at an SNR of 10 dB, the mean scores of SIG, BAK, and OVRL scales are also summarized in Tables 6, 7, and 8. The mean scores in the presence of both car and babble noises demonstrate that the lower signal distortion (i.e., higher SIG scores) and the lower noise distortion (i.e., higher BAK scores) are obtained with the proposed method relative to that obtained by Universal and SMPO methods in most of the conditions. It is also shown that a consistently better performance in OVRL scale is offered by the proposed method not only in car but also in babble noisy conditions at both SNR levels considered in comparison to that provided by all the methods mentioned above. Overall, it is found that the proposed method possesses the highest subjective sound quality in comparison to that of the other methods in case of different noises at various levels of SNRs.

## 4. Conclusions

In this paper, we developed a Gaussian statistical model-based technique for the TE operated PWP coefficients of the noisy speech in order to obtain a suitable threshold value. By employing the proposed gaussian pdf dependent custom thresholding function, the PWP coefficients of the noisy speech are thresholded in order to obtain an enhanced speech. Simulation results show that the proposed method yields consistently better results in the sense of higher Segmental SNR Improvement in dB, higher output PESQ, and lower WSS values than those of the existing methods.

Table 3: Mean Score of SIG scale for different methods in presence of car noise at 5 db

| Listener | Universal | SMPO | Proposed Method |
|---|---|---|---|
| 1 | 3.6 | 4.0 | 4.0 |
| 2 | 3.3 | 3.9 | 3.7 |
| 3 | 3.9 | 4.0 | 4.2 |
| 4 | 3.4 | 4.2 | 4.5 |
| 5 | 3.2 | 3.8 | 4.0 |
| 6 | 2.9 | 3.6 | 3.9 |
| 7 | 3.8 | 3.8 | 4.2 |
| 8 | 3.5 | 3.7 | 4.2 |
| 9 | 3.5 | 3.9 | 3.8 |
| 10 | 3.7 | 3.9 | 4.0 |

Table 4: Mean Score of BAK scale for different methods in presence of car noise at 5 db

| Listener | Universal | SMPO | Proposed Method |
|---|---|---|---|
| 1 | 4.0 | 4.5 | 5.0 |
| 2 | 4.3 | 4.9 | 4.7 |
| 3 | 4.2 | 4.4 | 4.9 |
| 4 | 4.4 | 4.7 | 4.8 |
| 5 | 4.2 | 4.8 | 4.7 |
| 6 | 3.9 | 4.6 | 4.9 |
| 7 | 3.8 | 3.9 | 4.4 |
| 8 | 4.4 | 4.6 | 4.6 |
| 9 | 3.5 | 3.8 | 4.5 |
| 10 | 4.2 | 4.5 | 4.8 |

Table 5: Mean Score of OVL scale for different methods in presence of car noise at 5 db

| Listener | Universal | SMPO | Proposed Method |
|---|---|---|---|
| 1 | 2.6 | 4.0 | 4.1 |
| 2 | 3.3 | 3.8 | 3.7 |
| 3 | 3.9 | 4.1 | 4.3 |
| 4 | 3.6 | 4.2 | 4.2 |
| 5 | 3.3 | 3.9 | 4.1 |
| 6 | 3.9 | 4.6 | 4.9 |
| 7 | 3.8 | 3.8 | 4.3 |
| 8 | 3.6 | 4.1 | 4.2 |
| 9 | 3.5 | 4.5 | 4.7 |
| 10 | 3.9 | 4.6 | 4.8 |

Table 6: Mean Score of SIG scale for different methods in presence of Babble noise at 5 db

| Listener | Universal | SMPO | Proposed Method |
|---|---|---|---|
| 1 | 3.6 | 4.0 | 4.0 |
| 2 | 3.3 | 3.9 | 3.7 |
| 3 | 4.2 | 3.9 | 4.0 |
| 4 | 3.4 | 4.2 | 4.5 |
| 5 | 3.2 | 3.8 | 4.0 |
| 6 | 2.9 | 3.6 | 3.9 |
| 7 | 3.8 | 3.8 | 4.2 |
| 8 | 3.4 | 3.6 | 4.1 |
| 9 | 3.5 | 3.9 | 3.7 |
| 10 | 3.7 | 3.8 | 3.9 |

Table 7: Mean Score of BAK scale for different methods in presence of Babble noise at 5 db

| Listener | Universal | SMPO | Proposed Method |
|---|---|---|---|
| 1 | 4.0 | 4.5 | 5.0 |
| 2 | 4.3 | 4.9 | 4.7 |
| 3 | 4.2 | 4.4 | 4.9 |
| 4 | 4.4 | 4.7 | 4.8 |
| 5 | 4.2 | 4.8 | 4.7 |
| 6 | 3.9 | 4.6 | 4.9 |
| 7 | 3.8 | 3.9 | 4.4 |
| 8 | 4.4 | 4.6 | 4.7 |
| 9 | 3.5 | 3.9 | 4.7 |
| 10 | 4.7 | 4.8 | 4.9 |

Table 8: Mean Score of OVL scale for different methods in presence of babble noise at 5 db

| Listener | Universal | SMPO | Proposed Method |
|---|---|---|---|
| 1 | 2.6 | 4.0 | 4.1 |
| 2 | 3.3 | 3.8 | 3.7 |
| 3 | 3.9 | 4.1 | 4.3 |
| 4 | 3.6 | 4.2 | 4.2 |
| 5 | 3.3 | 3.9 | 4.1 |
| 6 | 3.9 | 4.6 | 4.9 |
| 7 | 3.8 | 3.8 | 4.3 |
| 8 | 3.6 | 4.1 | 4.2 |
| 9 | 3.5 | 4.5 | 4.7 |
| 10 | 3.9 | 4.8 | 4.9 |

The improved performance of the proposed method is also indicated and attested by the much better spectrogram outputs and in terms of the higher scores in the formal subjective listening tests.